\documentclass[preprint,12pt,sort&compress]{elsarticle}

\usepackage{amssymb}

\usepackage{epsfig,amssymb,bm,graphics,color,amsmath}

\usepackage[]{units}
\usepackage{rotating, graphicx}
\usepackage{slashed}
\usepackage{chngcntr}
\usepackage{floatrow}
\usepackage{hyperref}
\definecolor{change}{rgb}{0.0, 0.0, 0.0}
\definecolor{change2}{rgb}{0.0, 0.0, 0.0} 

\counterwithout{footnote}{section}

\newcommand{\Tcepval}{112}
\newcommand{\mucepval}{612}
\newcommand{\TCEP}{$T_{CEP} \approx \Tcepval \text{ MeV}$}
\newcommand{\MUCEP}{$\mu_{CEP} \approx \mucepval \text{ MeV}$}

\journal{Physics Letters B}

\begin{document}

\begin{frontmatter}



\title{Holographic QCD phase diagram with critical point from Einstein-Maxwell-dilaton dynamics}


\author{J. Knaute}
\ead{j.knaute@hzdr.de}

\author{R. Yaresko, B. K\"ampfer}

\address{Helmholtz-Zentrum Dresden-Rossendorf,
POB 51 01 19, 01314 Dresden, Germany and\\
TU Dresden, Institut f\"ur Theoretische Physik, 01062 Dresden, Germany}

\begin{abstract}
Supplementing the holographic Einstein-Maxwell-dilaton model of 
\cite{DeWolfe:2010he,DeWolfe:2011ts} by input of lattice QCD data for 2+1 flavors and physical quark masses 
for the equation of state and quark number susceptibility at zero baryo-chemical potential 
we explore the resulting phase diagram over the temperature-chemical potential plane. 
A first-order phase transition sets in at a temperature of about $\Tcepval$ MeV and 
a baryo-chemical potential of {\color{change}$\mucepval$ MeV}. 
We estimate the accuracy of the critical point position in the order of approximately {\color{change}$5-8\,\%$ by considering parameter variations and} different low-temperature asymptotics for the second-order quark number susceptibility. 
The critical pressure as a function of the temperature has a positive slope, i.e.\ the entropy per baryon jumps up when crossing the phase border line from larger values of temperature/baryo-chemical potential, thus classifying the phase transition as a gas-liquid one. The updated holographic model exhibits in- and outgoing isentropes in the vicinity of the first-order phase transition. 
\end{abstract}

\begin{keyword}
gravity dual \sep holography \sep quark-gluon plasma \sep critical point
\PACS 11.25.Tq \sep 47.17.+e \sep 05.70.Ce \sep 12.38.Mh \sep 21.65.Mn
\end{keyword}

\end{frontmatter}


\section{Introduction}

The QCD phase diagram exhibits potentially a large variety of structures \cite{Fukushima:2013rx,Ding:2015ona,Fukushima:2015bda,Braun-Munzinger:2015hba}. 
Either originating from extrapolations of weak-coupling results or being 
suggested by models (most notably Nambu--Jona-Lasinio (cf.\ \cite{Ranea-Sandoval:2015ldr}), linear sigma/quark-meson \cite{Olbrich:2015gln}
models in numerous variants), 
various phases of strongly interacting matter may occur, 
such as color superconductors (cf.\ \cite{Rapp:1997zu,Schafer:1999jg,Alford:2007xm}), or quarkyonic matter (cf.\ \cite{Andronic:2009gj}), 
or chirally restored phases (cf.\ \cite{Eser:2015pka}), or color-flavor locked structures (cf.\ \cite{Nickel:2006kc}).

While the gas-liquid (GL) first-order phase transition (FOPT) in nuclear matter 
seems to be well established since some time \cite{Finn:1982tc,Minich:1982tb,Hirsch:1984yj,Pochodzalla:1995xy,Natowitz:2002nw,Karnaukhov:2003vp}, 
the hadron-quark (HQ) deconfinement transition still offers a few challenges. 
At very small or zero net-baryon density corresponding to a small chemical potential 
($\mu$)-to-temperature ($T$) ratio, $\mu/T \ll 1$, the HQ transition is established as a
crossover in 2+1 flavor lattice QCD with physical quark masses \cite{Borsanyi:2013bia,Bazavov:2014pvz} at 
a characteristic scale of $T_c = \mathcal{O}(150 \text{ MeV})$. 
The popular Columbia plot \cite{Brown:1990ev} sketches qualitatively the options 
of the phase structure in dependence of the u, d, s quark masses $m_{u,d,s}$. 
For instance, in the chiral limit, $m_{u,d,s} \rightarrow 0$, or the opposite
infinitely heavy quark-mass limit, $m_{u,d,s} \rightarrow \infty$, the deconfinement 
transition is a FOPT. Due to the sign problem of the fermionic determinant the 
ab initio lattice QCD evaluations are not yet conclusive with respect to the confinement
and chiral restoration transition(s) at non-zero baryo-chemical potential, in particular for $\mu/T>2$. 
Some methods try to avoid or circumvent the sign problem (cf.\ \cite{Stephanov:2007fk}), 
e.g.\ by evaluations at imaginary $\mu$ (which need a prescription of $i\mu \rightarrow \mu$) or 
a Taylor expansion in powers of $\mu/T$ with coefficients calculated at $\mu = 0$
(which needs statements on the convergence \cite{Bazavov:2017dus}), 
or the reweighting method (which needs statements on the density and parameter ranges to incorporate the sign and overlap problem~\cite{Iwami:2015eba}).
{\color{change}Other approaches are based on the complex Langevin method \cite{Parisi:1980ys,Parisi:1984cs} 
(see \cite{Aarts:2016qhx} for recent developments) or a recent proposal for a path optimization method 
\cite{Mori:2017pne}, which is based on the Lefschetz-thimble path-integral method \cite{Cristoforetti:2012su}.}

The pertinent uncertainties make the region of larger $\mu/T$ interesting. 
A particularly interesting option is the possibility of a (critical) end point (CEP) 
of a curve of FOPTs, e.g.\ $T_c(\mu)$, setting in at $(T_{CEP}, \mu_{CEP})$ and 
running toward the $T = 0$ axis when imaging the phase diagram in the $T\!-\!\mu$ plane.

The CEP coordinates are yet fairly unconstrained. Plugging model results 
and QCD-related extrapolations together one arrives at some less conclusive 
scatter plot (cf.\ e.g.\ \cite{Stephanov:2007fk}). Advanced lattice QCD approaches disfavor a CEP position at $T/T_c(\mu=0)>0.9$ and $\mu/T \le 2$ \cite{Bazavov:2017dus}.

Experimentally, there are dedicated programs aiming at pinning down the CEP location. 
For instance, the beam energy scan at RHIC \cite{Luo:2015ewa} gave hints on some features in the beam
energy dependence of selected observables which have been interpreted as CEP signature
(cf.\ \cite{Mustafa:2015yeg}). In \cite{Lacey:2015yxg} another view has been launched with the conclusion of 
having also seen CEP indications.
Furthermore, the SHINE (NA61) collaboration at CERN-SPS is also systematically 
seeking CEP effects \cite{Andronov:2015ucu}. Experiments planned at FAIR and NICA and J-PARC \cite{Sako:2014fha} are 
analogously driven by CEP searches, analoguously as goals by the CBM collaboration \cite{cbm_web,Friman:2011zz}, and the MPD group \cite{Sorin:2011zz}. 

Given that challenges from both theory and experiment one can ask whether 
further theoretical model classes beyond the above mentioned approaches could 
be useful in exploring the hypothetical FOPT emerging from a CEP. Holographic
models, advancing the seminal AdS/CFT correspondence \cite{Maldacena:1997re,Gubser:1998bc,Witten:1998qj}, are thought to 
mimic essential QCD properties in the strong-coupling regime
\cite{Schafer:2009dj,CasalderreySolana:2011us,Adams:2012th,DeWolfe:2013cua,Brambilla:2014jmp} and thus may serve as suitable candidates for such an enterprise. 
{\color{change2}Holographic bottom-up approaches coupled to a self-interacting dilaton with nontrivial potential were particularly successful to describe nonconformal properties of the quark-gluon plasma and QCD \cite{Csaki:2006ji,Gursoy:2007cb,Gursoy:2007er,Gubser:2008ny}.}
In \cite{DeWolfe:2010he,DeWolfe:2011ts} a model
formulation has been put forward which displays a critical point in the 
$T\!\!-\!\!\mu$ plane. While \cite{DeWolfe:2010he,DeWolfe:2011ts} focuses on CEP properties and an outline of 
some transport coefficients, \cite{Rougemont:2015wca,Rougemont:2015ona} employed that holographic model 
to investigate thermodynamics and further transport quantities at small $\mu/T$, however, the question of the CEP position, based on an adjustment to  recent lattice data, and properties of phase diagrams were not addressed.
The model rests on the coupled Einstein-Maxwell-dilaton (EMd) dynamics and 
can be adjusted to QCD thermodynamics, i.e.\ the equation of state (EoS) and 
quark number susceptibility at $\mu = 0$. The resulting phase structure is 
the topic of our present paper. We feel that an update of \cite{DeWolfe:2010he,DeWolfe:2011ts} is timely since by now consistent and more precise lattice QCD data are at our disposal. In fact, we find some some qualitatively important modifications in comparison to \cite{DeWolfe:2010he,DeWolfe:2011ts} w.r.t.\ the pattern of isentropes in the phase diagrams as well as the position of the CEP.

With respect to the discussion in \cite{Steinheimer:2013xxa}, a FOPT curve 
is specified by further peculiarities: it can be related either to a GL 
type or to a HQ type transition.
For a discussion contrasting features of GL and HQ phase transitions we refer 
the interested reader to \cite{Steinheimer:2013xxa,Iosilevskiy:2015sia,Hempel:2015eoj,Wunderlich:2016aed},
where the notions of entropic vs.\ enthalpic transitions as well as congruent 
and non-congruent material changes are exemplified and representations in other 
variables than $T\!-\!\mu$ are exhibited.
Such different FOPTs can matter significantly in core-collapse 
supernova explosions as discussed in some detail in \cite{Hempel:2015vlg}. 
Motivated by such a relation to astrophysical aspects of the phase structure 
of strongly interacting matter~- not only touching core-collapse dynamics 
but also neutron (quark core) stars~- we unravel here the phase structure 
of the holographic EMd model. It turns out that the EMd model with adjustments 
to QCD input belongs to the GL class. That is across the phase boundary both 
the baryon density $n$ and the entropy density $s$ jump when considering the stable phases. 
For the GL transition, the entropy per baryon $s/n$ drops down when going into $\mu$ 
or $T$ direction, while at the HQ transition $s/n$ jumps up, according to 
expectations in \cite{Steinheimer:2013xxa}. According to the Clausius-Clapeyron 
equation one finds the critical pressure $p(T, \mu_c(T))$ either with 
positive slope (GL transition) or with negative slope (HQ transition).\,\footnote{Obviously, 
the resulting behavior of the pressure at the FOPT at smaller 
temperatures is markedly depending on these details, with impact on the stiffness
of the EoS which in turn governs the possibility of a third family of compact 
stars or twin configurations \cite{Kampfer:1985mre,Schertler:2000xq,Alford:2015dpa,Alvarez-Castillo:2016oln}, 
on which the options for core-collapse supernova explosions according to 
\cite{Hempel:2015vlg} (and further references therein) depend on.}

Our paper is organized as follows. In section \ref{sec:EMdmodel} we recall the holographic EMd model.
The numerical adjustment to lattice QCD data at $\mu=0$ is described in section \ref{sec:mu0} and the numerical results for the phase diagrams are presented in section \ref{sec:PD}, including an analysis of  the impact of different assumptions for the susceptibility at small temperatures. We summarize in section \ref{sec:summary}.

\section{Recalling the holographic EMd model}
\label{sec:EMdmodel}

The holographic model of gravity of a 5-dimensional Riemann space sourced by the 
coupled Maxwell-dilaton fields is defined in \cite{DeWolfe:2010he,DeWolfe:2011ts} by the action
 \begin{equation} \label{eq:S}
S = \frac{1}{2 \kappa_{5}^2}\int d^{5}x\sqrt{-g}\left(R-\frac{1}{2}\partial^{\mu}\phi\partial_{\mu}\phi-V(\phi) - \frac{f(\phi)}{4}F^2_{\mu\nu}\right) + S_{GH}, 
\end{equation}
where $R$ is the Einstein-Hilbert part, $F_{\mu\nu} = \partial_\mu A_{\nu} - \partial_\nu A_{\mu}$ with  
$A_\mu dx^\mu = \Phi dt$ stands for the Abelian 
gauge field \`a la Maxwell, and $\phi$ is a real scalar (dilaton) with 
self-interaction described by the so called potential $V(\phi)$. 
The Maxwell field and dilaton are coupled by a dynamical strength function $f(\phi)$. 
The Gibbons-Hawking term $S_{GH}$ for a consistent formulation of the variational problem is not 
needed explicitly in our context. 
The ``Einstein constant'' $\kappa_5$ is taken as a model parameter.
The ansatz for the infinitesimal line 
element squared
\begin{equation} \label{eq:ds}
 ds^2 = e^{2 A(r; r_H)} \big( - h(r; r_H) dt^2 + d \vec x^2 \big) + \frac{e^{2B(r; r_H)}dr^2}{h(r; r_H)}
\end{equation} 
highlights that (i) only the dynamics in bulk direction $r$ is considered and 
(ii) a horizon is admitted at $r = r_H$ by a simple zero of the blackness function $h$. 
By a gauge choice, one can achieve $B = 0$ and $r_H = 0$. 
We solve the field equations following from (\ref{eq:S}, \ref{eq:ds}) with the technique described in \cite{DeWolfe:2010he,DeWolfe:2011ts}. 
In a nutshell:
One has to numerically integrate from $r_H + \epsilon$ towards the boundary at $r \to \infty$.
Requiring regularity of $A, h, \phi, \Phi$ at the horizon $r = r_H$, defined by $h(r_H; r_H) = 0$,
series solutions for any these functions can be obtained, which yield the initial conditions
for the integration.
After fixing all gauge redundancies the
two remaining independent quantities parametrizing the solutions are $\phi_0 \equiv \phi(r_H, r_H)$
and $\Phi_1 \equiv \frac{\partial \Phi}{\partial r}\big\vert_{r_H}$. 
It follows from the horizon expansion of $A$
that $\Phi_1$ is bounded, $\Phi_1 < \Phi_1^{max} \equiv \sqrt{- \frac{2V(\phi_0)}{f(\phi_0)}}$.
Close to the boundary, the following expansions 
in powers of $e^{-\alpha(r)} \equiv \exp[- \frac{r}{L \sqrt{h_0^{\infty}}} - A_0^{\infty}]$
are valid:
$h(r) = h_0^{\infty} + \ldots$, 
$A(r) = \alpha(r) + \ldots$,
$\Phi(r) = \Phi_0^{\infty} + \Phi_2^{\infty} e^{-2 \alpha(r)} + \ldots$, and
$\phi(r) = \phi_A e^{ - (4 - \Delta) \alpha(r)}  + \phi_B e^{- \Delta \alpha(r)} + \ldots$.
The expansion of $\phi$ assumes 
$L^2 V(\phi) = - 12 + \frac 12 [\Delta(\Delta - 4)] \phi^2 + \ldots$ for $\phi \rightarrow 0$.\,\footnote{This means 
we are considering a relevant operator in the boundary theory with scaling dimension 
$\Delta < 4$; see \cite{Gursoy:2008za} for a 
different choice of potential asymptotics that correspond to a marginal operator.}
By the standard AdS/CFT dictionary, 
$\phi_A$ is the source and $\phi_B$ the expectation value of the boundary theory operator dual to $\phi$.
Then one obtains the thermodynamic quantities 
temperature $T$, entropy density $s$, baryo-chemical potential $\mu$ and 
baryon density $n$ as
\begin{eqnarray}
  T &=& \lambda_T          \frac{1}{4\pi \phi_A^{1/(4 - \Delta)} \sqrt{h_0^{\infty}}}, \label{eq:T} \\
  s &=& \lambda_s          \frac{2 \pi}{\phi_A^{3/(4 - \Delta)}}, \label{eq:s} \\
  \mu &=& \lambda_\mu  \color{change}\frac{\Phi_0^{\infty}}{\phi_A^{1/(4 - \Delta)} \sqrt{h_0^\infty}}, \label{eq:mu} \\
  n &=& \lambda_n          \color{change}\frac{Q_G}{2 f(0) \phi_A^{3/(4 - \Delta)}} . \label{eq:n}
\end{eqnarray}
The dimensional scaling factors $\lambda_{T,s,\mu,n}$ are introduced as in \cite{DeWolfe:2010he,DeWolfe:2011ts} 
to compensate the arbitrary  choice $\kappa_5=L=1$ at intermediate steps and restore afterwards the physical units 
(here, $L$ is the AdS scale).
At the horizon, the radially conserved Gauss charge becomes $Q_G = f(\phi_0) \Phi_1$.
In such a way one maps out the $T\!-\!\mu$ plane by 
suitably chosen pairs $(\phi_0, \Phi_1)$ which entirely parametrize initial conditions at $r_H$; \eqref{eq:s} and \eqref{eq:n} deliver in 
each point $s$ and $n$. The pressure follows from the integration of 
$d p(T, \mu) = s(T, \mu) dT + n(T, \mu) d \mu$,
with $p(0, 0) = 0$, where $T= 0, \mu = 0$ correspond to $\phi_0 \rightarrow \infty, \Phi_1 = 0$. \\
The given bottom-up approach is to be supplemented by fixing the dilaton potential 
$V(\phi)$ and the dynamical coupling $f(\phi)$, e.g.\ from lattice QCD results
at $\mu = 0$. By properly engineering $V(\phi)$ one essentially dials the EoS at $\mu = 0$,
in particular whether a FOPT is built in (as for pure glue dynamics 
or QCD in the chiral limit(s)) or a crossover is incorporated (as for 2+1 flavor 
QCD with physical quark masses), see \cite{Yaresko:2013tia,Yaresko:2015ysa} for recent examples and 
\cite{Gubser:2008ny,Gursoy:2008za} for full-fledged pioneering investigations. Adjusting $f(\phi)$ 
at the quark number susceptibility from lattice QCD at $\mu = 0$ completes the model. 

Beyond $p-s-n$ thermodynamics also fluctuation measures, such as susceptibilities, variance, 
kurtosis, etc.\ follow via derivatives of the pressure. The susceptibilities are defined, 
in general, by 
$\chi_{i}(T, \mu) \equiv \frac{\partial^{i} p(T, \mu)}{\partial \mu^{i}}\Big\vert_T$, $i = 2,3,4,\ldots$.
By CP invariance, odd susceptibilities $\chi_{3,5,\ldots}$ at $\mu = 0$ vanish. 
In \cite{DeWolfe:2010he,DeWolfe:2011ts}, a comfortable formula for $\chi_2$ at $\mu = 0$ is given (see also \cite{Rougemont:2015wca,Rougemont:2015ona})
\begin{equation}
 \frac{\chi_2(T, 0)}{T^2} = \frac{L}{16\pi^2 f(0)} \frac{s}{T^3} \frac{1}{\int_{r_H}^\infty dr \frac{e^{-2A(r)}}{f(\phi(r))}},  
 \label{eqn:chi2T2}
\end{equation}
which allows the matching of $f(\phi)$ to lattice data. 

The EMd model (\ref{eq:S}, \ref{eq:ds}) with these input data is then ready to transport 
the information from $\mu = 0$ to $\mu > 0$, up to $T = 0$, thus uncovering the $T\!-\!\mu$ plane. 
This is very much the spirit of the quasi-particle model \cite{Peshier:1999ww,Peshier:2002ww,Plumari:2011mk}, 
where a flow equation facilitates such a transport.

\section{Adjustment to lattice QCD data at $\mu=0$}
\label{sec:mu0}

\begin{figure}[!t]
\centering
 \includegraphics[width=0.495\textwidth]{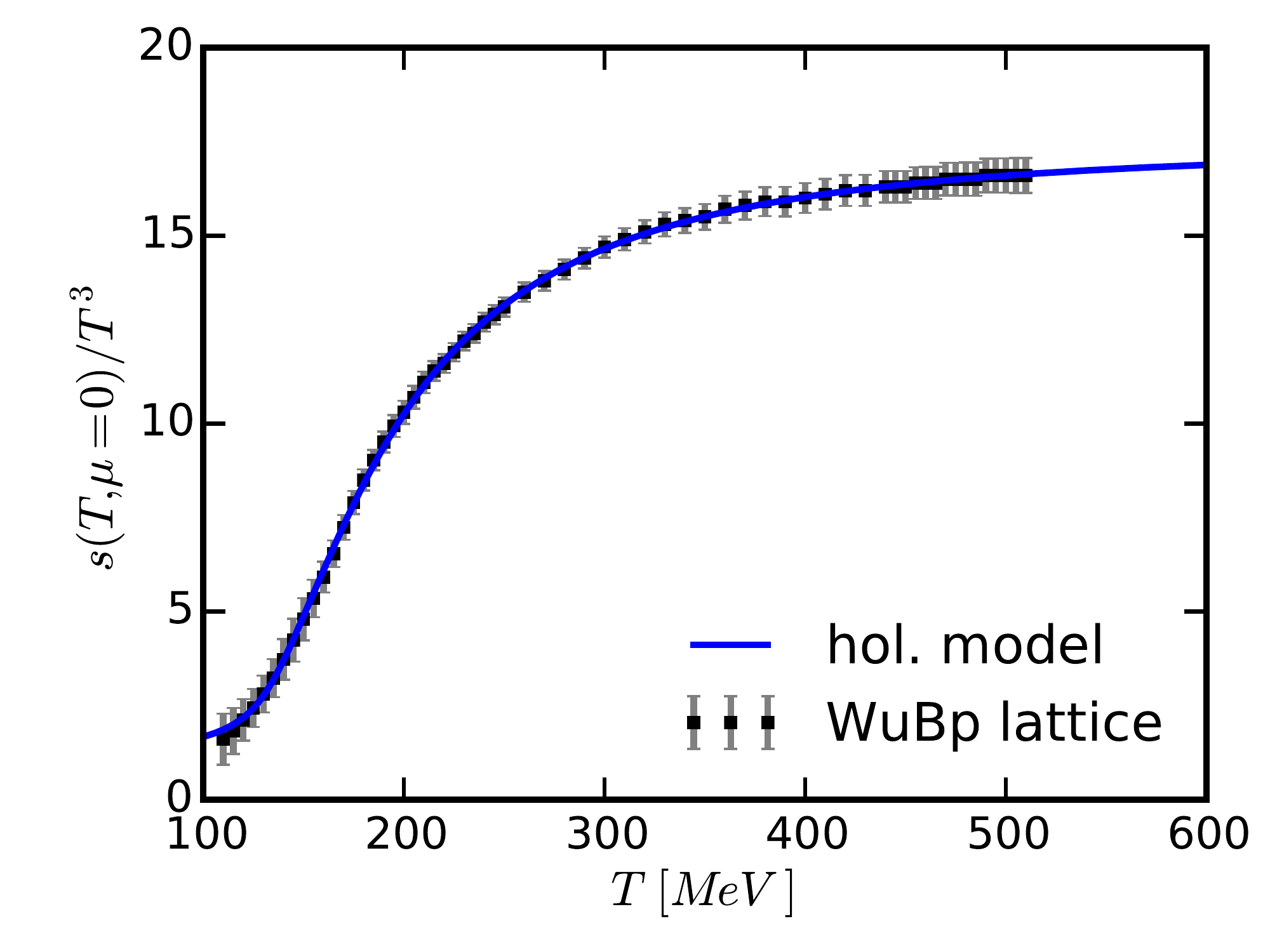}
 \includegraphics[width=0.495\textwidth]{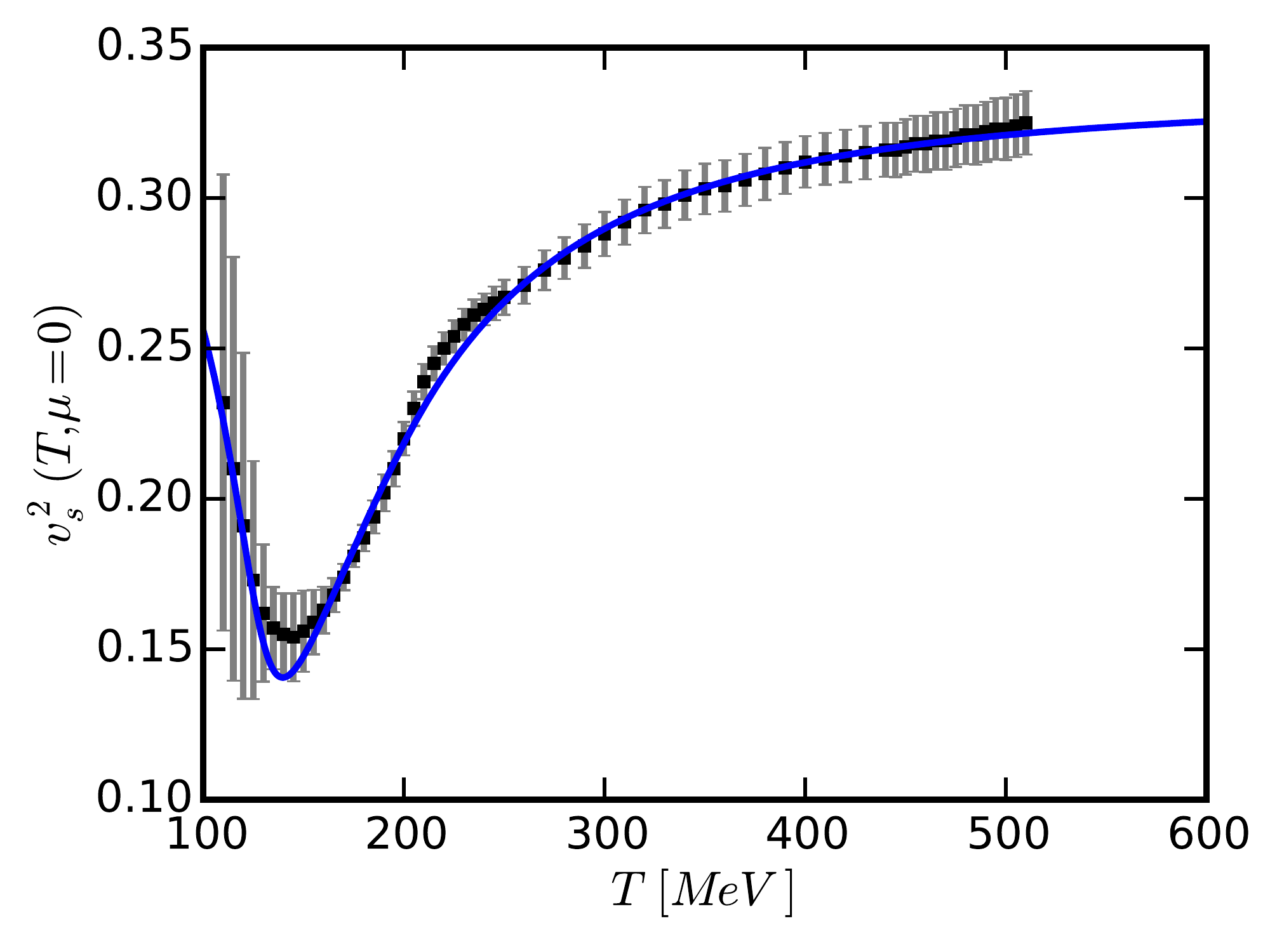}
 \includegraphics[width=0.495\textwidth]{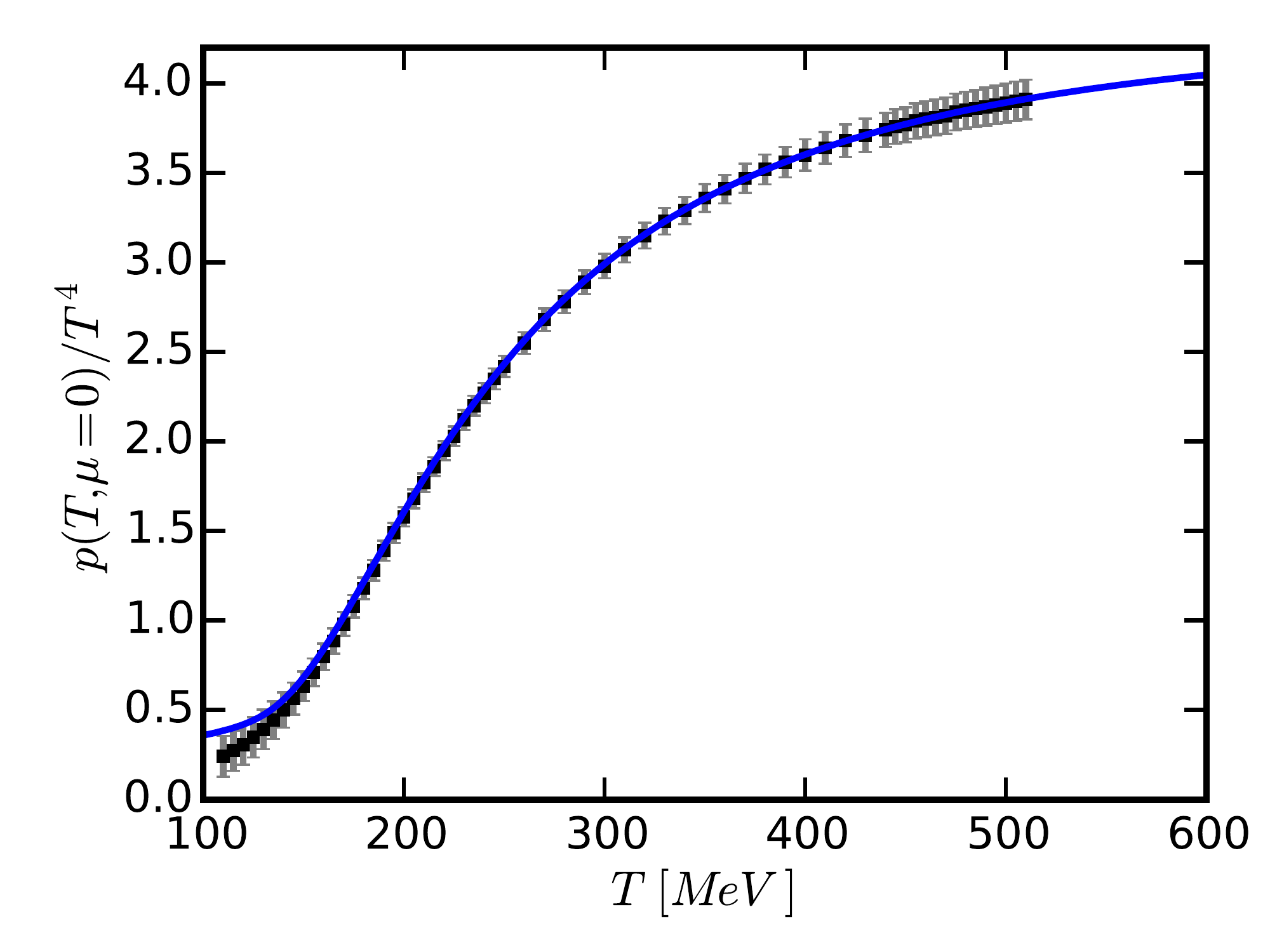}
 \includegraphics[width=0.495\textwidth]{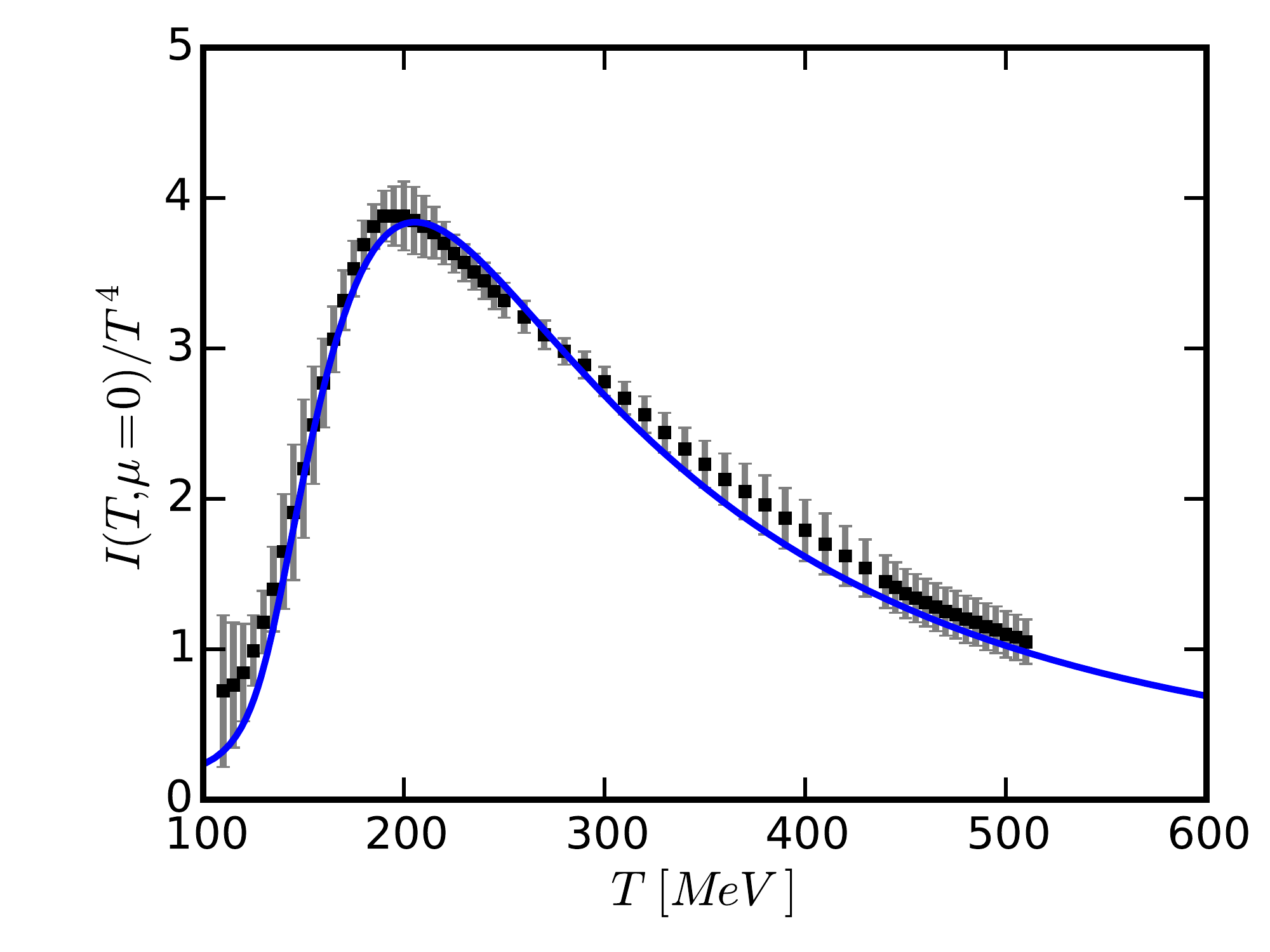}
 \caption{Equation of state of the updated holographic EMd model with parametrizations (\ref{eqn:V})-(\ref{eqn:lambdasparam}) as functions of $T$ for $\mu = 0$: scaled entropy density (top left), speed of sound squared $v_s^2=\frac{\partial\log T}{\partial\log s}$ (top right), scaled pressure (bottom left) and scaled trace anomaly (bottom right). Lattice results from \cite{Borsanyi:2013bia} are displayed as symbols with error bars.}
 \label{fig:EOS}
\end{figure}

\begin{figure}[!t]
\centering
 \includegraphics[width=0.495\textwidth]{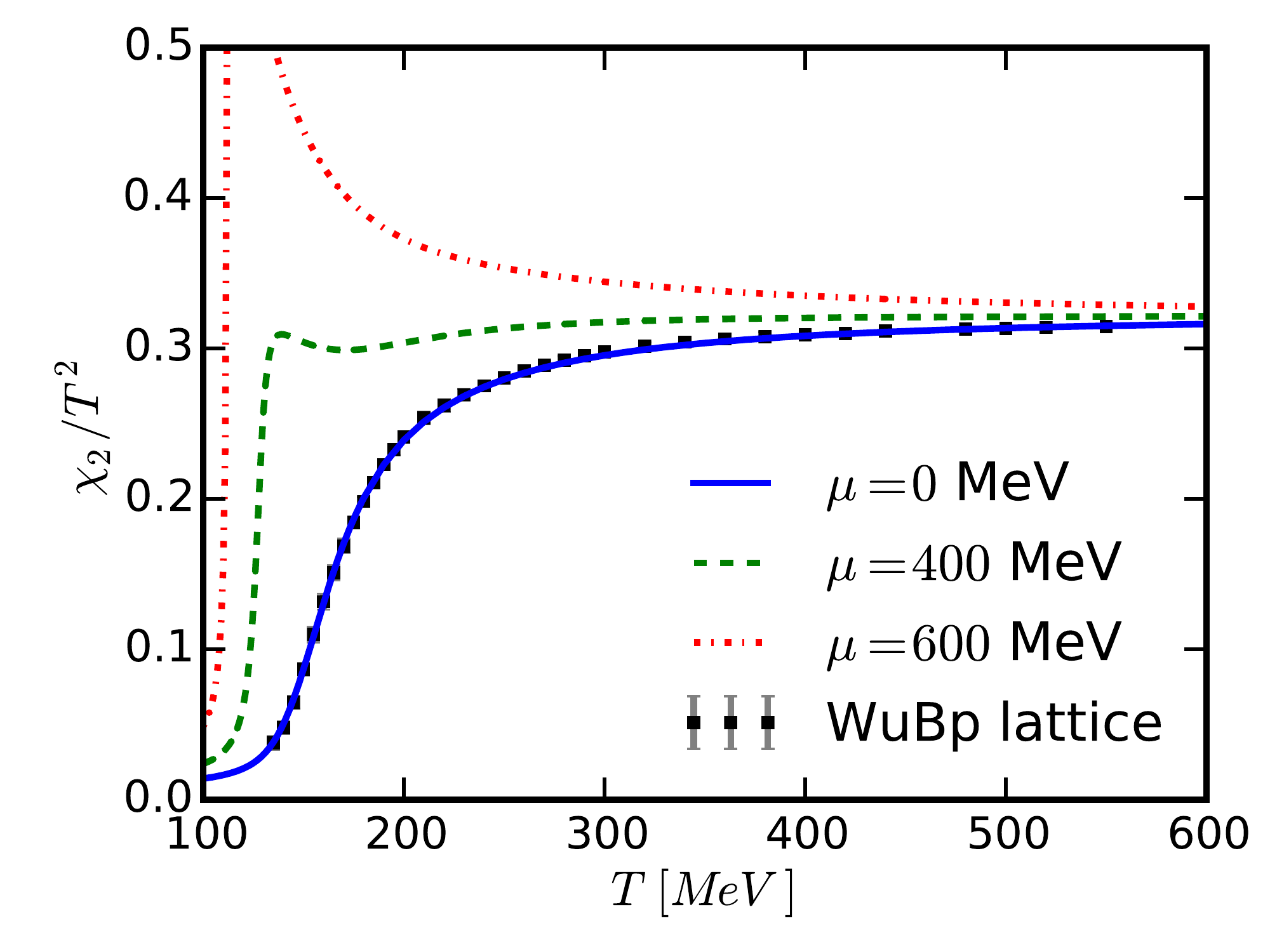}
 \includegraphics[width=0.495\textwidth]{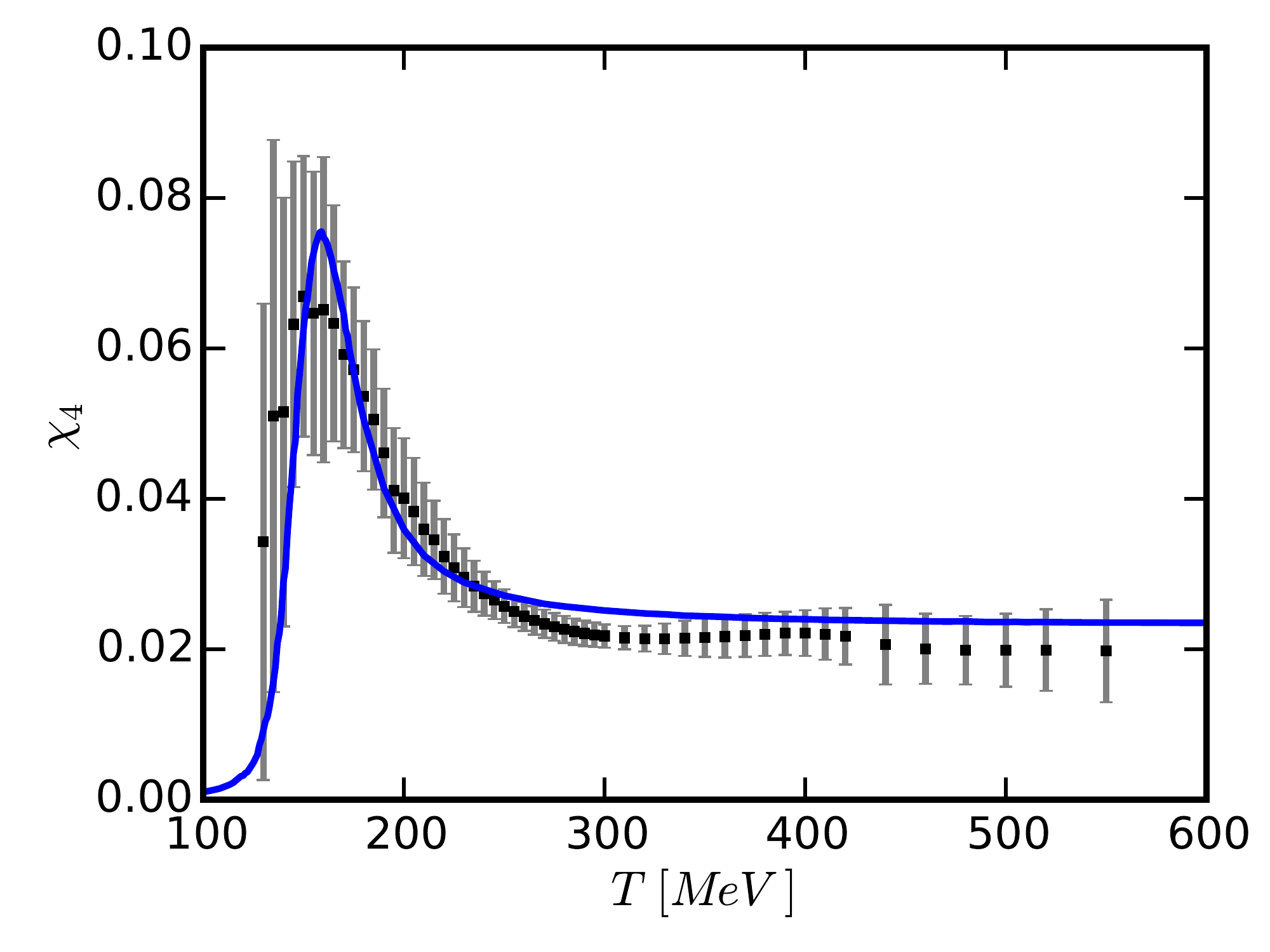}
\caption{Second-order susceptibility $\chi_2/T^2$ (left panel) {\color{change}and fourth-order susceptibility $\chi_4$ (right panel)} of the updated holographic EMd model with parametrizations (\ref{eqn:V})-(\ref{eqn:lambdasparam}) as function of $T$ {\color{change}for different values of the chemical potential $\mu$}. Lattice results from \cite{Bellwied:2015lba} are displayed as symbols with error bars.}
 \label{fig:chis}
\end{figure}

Contrary to \cite{Rougemont:2015wca,Rougemont:2015ona} we rely here on a modified previous fit \cite{Yaresko:2015ysa} to the 
2+1 flavor QCD thermodynamics with physical quark masses \cite{Borsanyi:2013bia,Bazavov:2014pvz} by 
the dilaton potential 
\begin{alignat}{1}
L^2 V(\phi) = 
 \begin{cases} -12 \exp\left\{\frac{a_1}{2}\phi^2 + \frac{a_2}{4}\phi^4 \right\} &: \phi < \phi_m\\ 
 a_{10} \cosh\left[ a_4(\phi-a_5)\right]^{a_3/a_4} \exp\left\{ a_6\phi + \frac{a_7}{a_8}\tanh\left[ a_8(\phi-a_9) \right] \right\}\mkern-14mu &: \phi \geq \phi_m
\end{cases} 
\label{eqn:V}
\end{alignat}
with parameters
\begin{equation}
\begin{split}
\begin{tabular}{c|c|c|c|c|c}
$\phi_m$ & $a_1$ & $a_2$ & $a_3$ & $a_4$ & $a_5$   \\ \hline 
1.7058 & 0.2840 & -0.0089 & 0.7065 & 0.4951 & 0.1761
\end{tabular} \ , \\
\begin{tabular}{c|c|c|c|c}
 $a_6$ & $a_7$ & $a_8$ & $a_9$ & $a_{10}$  \\ \hline 
 -0.0113 & -0.4701 & 2.1420 & 4.3150 & -10.0138
\end{tabular} \ ,
\label{eqn:Vparam}
\end{split}
\end{equation}
implying $\Delta=2(1+\sqrt{1-3 a_1})$. 
A fit of $\chi_2/T^2$ to data in \cite{Bellwied:2015lba} by the ansatz in \eqref{eqn:chi2T2} 
\begin{equation}
 f(\phi) = c_0 + c_1\tanh\left[ c_2(\phi-c_3)\right] \color{change}+ c_4 \exp\left[-c_5 \phi\right]
 \label{eqn:f}
\end{equation}
delivers the parameters 
\begin{equation}
\begin{tabular}{c|c|c|c|c|c}
$c_0$ & $c_1$ & $c_2$ & $c_3$ & $\color{change}c_4$ & $\color{change}c_5$ \\ \hline 
0.1892 & -0.1659 & 1.5497 & 2.1820 & \color{change}0.6219 & \color{change}112.7136 
\end{tabular} 
 \label{eqn:fparam}
\end{equation}
together with the fit results\,\footnote{\color{change}In \cite{Knaute:2017opk} we allowed for four independent scale setting
parameters $\lambda_{T,s,\mu,n}$ as in \cite{DeWolfe:2010he,DeWolfe:2011ts} and a different ansatz for 
$f(\phi)$, resulting in different values for quantities referring to the $\mu$ dependence.
Here, we enforce $\lambda_T = \lambda_\mu$ and $\lambda_s = \lambda_n$, as in \cite{Rougemont:2015wca,Rougemont:2015ona}, to accommodate the only two scales encoded in $\kappa_5$ and $L$.}
\begin{equation}
\begin{tabular}{c|c}
$\color{change}\lambda_T = \lambda_\mu$ & $\color{change}\lambda_s = \lambda_n$   \\ \hline 
$\unit[1148.07]{MeV}$ & $(\unit[513.01]{MeV})^3$ 
\end{tabular} \ .
\label{eqn:lambdasparam}
\end{equation}
Uncharged black hole solutions are numerically generated with initial conditions $\Phi_1=0$ and $\phi_0 \in [0.35, 5.0]$.
The resulting equation of state for $\mu=0$ is shown in Fig.\,\ref{fig:EOS} and the corresponding susceptibilities are exhibited in Fig.\,\ref{fig:chis} (blue curves).\,\footnote{\color{change}The fourth-order susceptibility $\chi_4$ is calculated using smoothed spline derivatives w.r.t.\ $n(\mu, T=\text{const})$. 
We verified the robustness of this numerical procedure for different smoothing conditions. Results for $\chi_2/T^2$ at finite $\mu$ are obtained similarly without smoothing technique.} 
We emphasize the consistency of the lattice data in \cite{Borsanyi:2013bia} and \cite{Bazavov:2014pvz} and select the data of \cite{Borsanyi:2013bia} for a comparison. The multi-parameter ans\"atze (\ref{eqn:V})-(\ref{eqn:lambdasparam}) allow in fact a fairly precise description of the available data.\,\footnote{\color{change}The parameters
(\ref{eqn:Vparam}), (\ref{eqn:fparam}) and (\ref{eqn:lambdasparam}) represent our best fit values and are used for the following studies. 
Variations in the order of $0.8\,\%$ of the potential parameters (\ref{eqn:Vparam}) and $3\,\%$ of the coupling function parameters (\ref{eqn:fparam}) still allow a good description within the uncertainties of the lattice results.}

\section{Phase diagram}
\label{sec:PD}

\begin{figure}[!t]
\centering
  \includegraphics[trim=0 0 110 0, clip, width=0.495\textwidth]{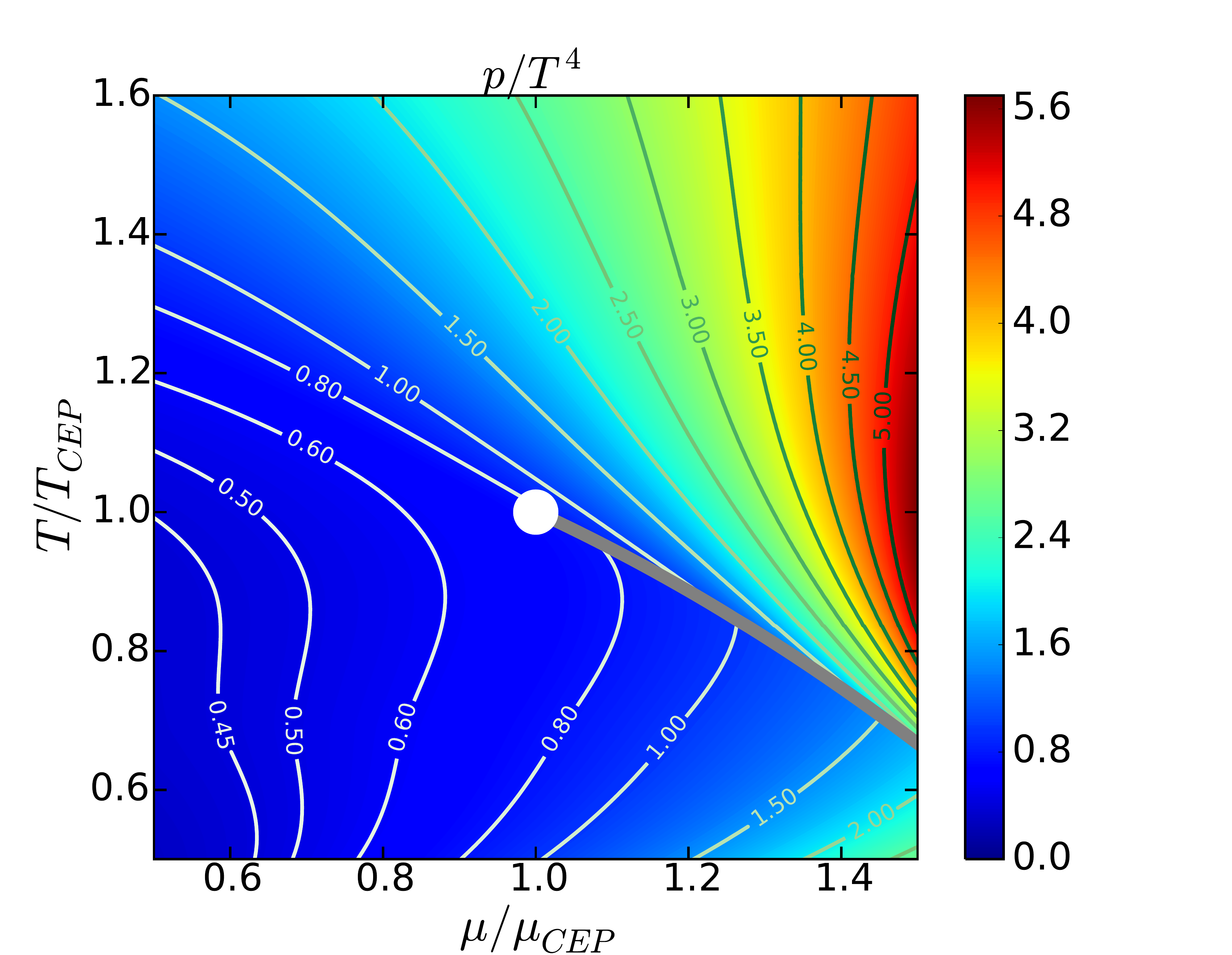}
  \includegraphics[trim=0 0 110 0, clip, width=0.495\textwidth]{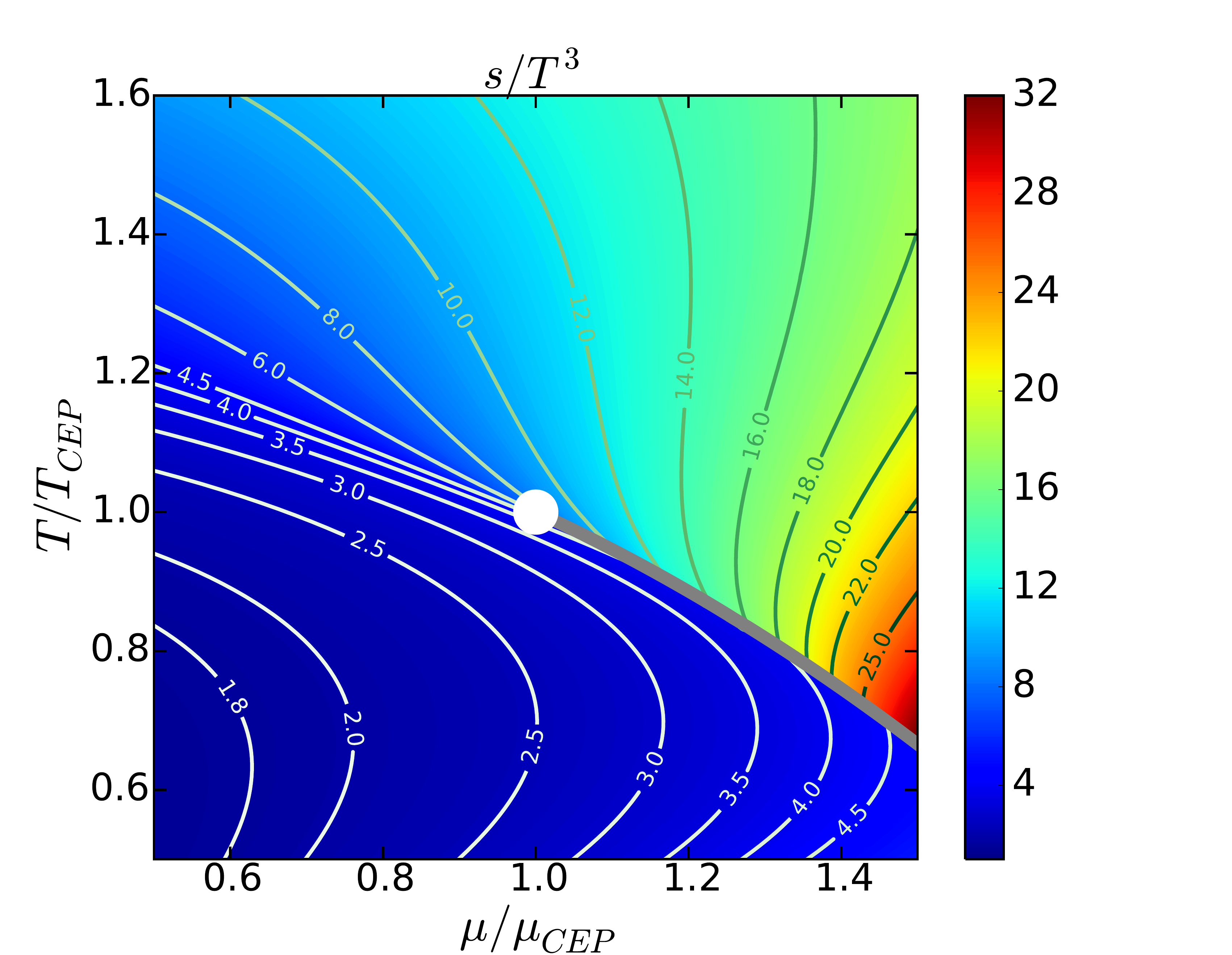} 
  \includegraphics[trim=0 0 110 0, clip, width=0.495\textwidth]{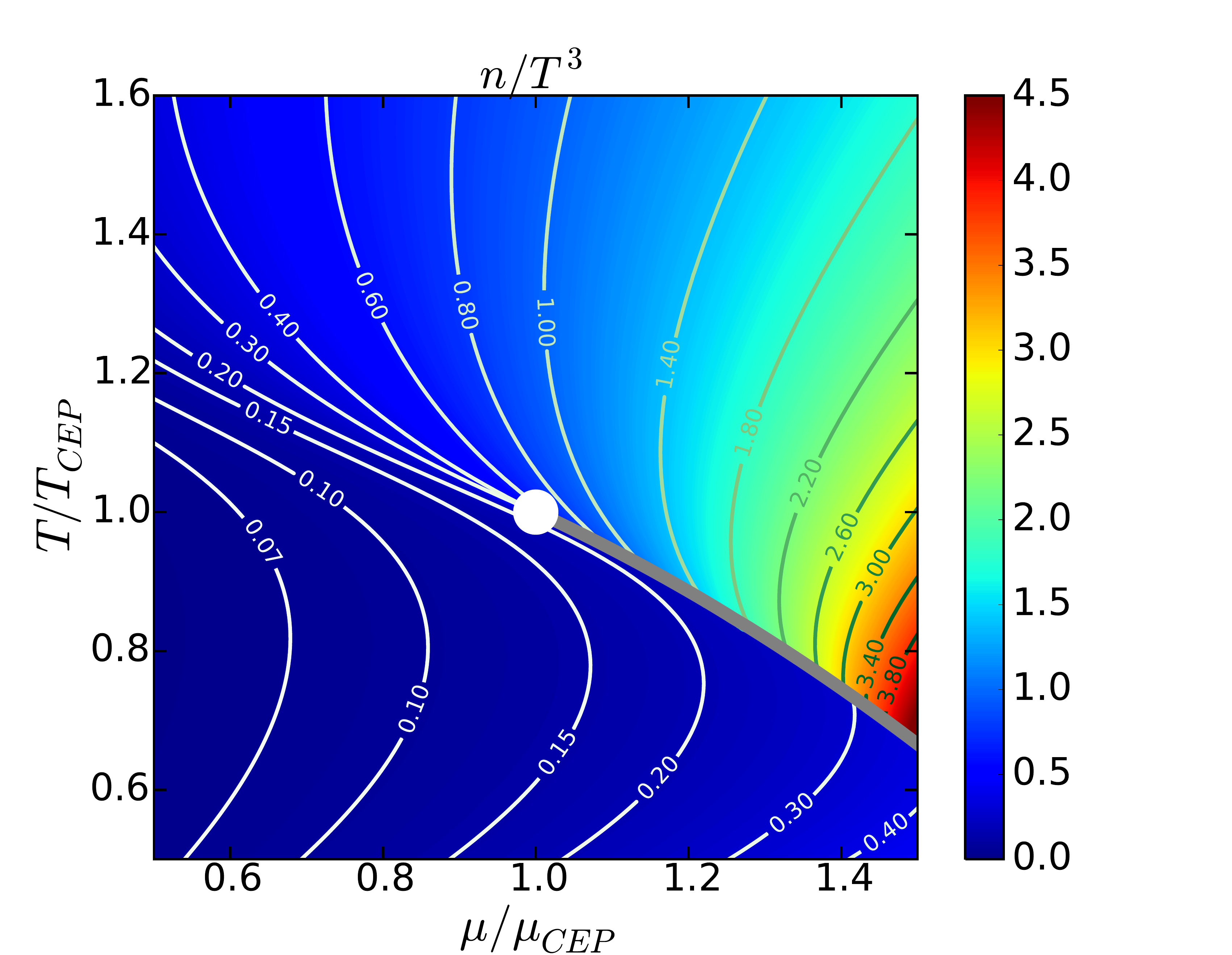}
  \includegraphics[trim=0 0 110 0, clip, width=0.495\textwidth]{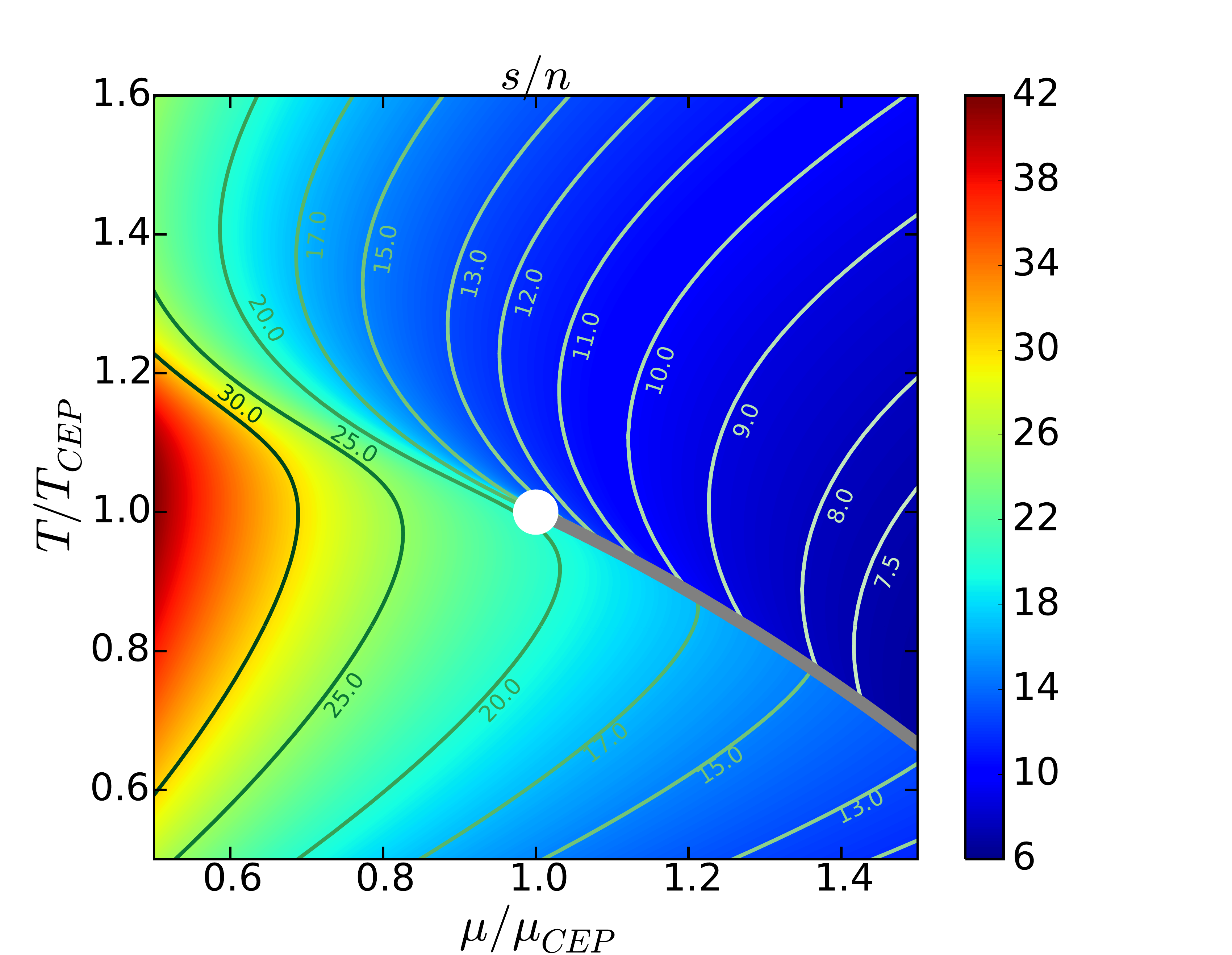} 
 \caption{Contour plots of scaled pressure (top left), scaled entropy density (top right), scaled baryon density (bottom left), and entropy-to-baryon ratio (bottom right) over the scaled $T-\mu$ plane 
 for the updated holographic EMd model. The position of the CEP is marked by a white dot and the FOPT curve is displayed as grey line.}
 \label{fig:PD}
\end{figure}

Charged black hole solutions with initial conditions $\phi_0 \in [{\color{change}0.35}, 4.5]$ and \linebreak$\Phi_1/\Phi_1^{max}(\phi_0) \in [0, {\color{change}0.755}]$ result in the thermodynamic phase diagram exhibited in Fig.\,\ref{fig:PD} in various variants over the $T\!-\!\mu$ plane. 
Only the stable-phase quantities are shown, i.e.\ in the case of multi-valued solutions at a given $T\!-\!\mu$ point those with maximum pressure. 
The CEP coordinates are {\color{change}$T_{CEP}=\unit[(111.5\pm0.5)]{MeV}$ and $\mu_{CEP}=\unit[(611.5\pm0.5)]{MeV}$}.
{\color{change}These uncertainties are estimated through the numerical calculation of discrete levels of constant temperature and chemical potential.}\,\footnote{\color{change}The significantly larger value of $\mu_{CEP}$ reported in \cite{Knaute:2017opk} is a consequence of a rough scaling with $\lambda_\mu/\lambda_T$ if both $\lambda_T$ and $\lambda_\mu$ are unconstrained.}

\begin{figure}[!t]
\centering
 \includegraphics[trim=0 0 110 0, clip, width=0.495\textwidth]{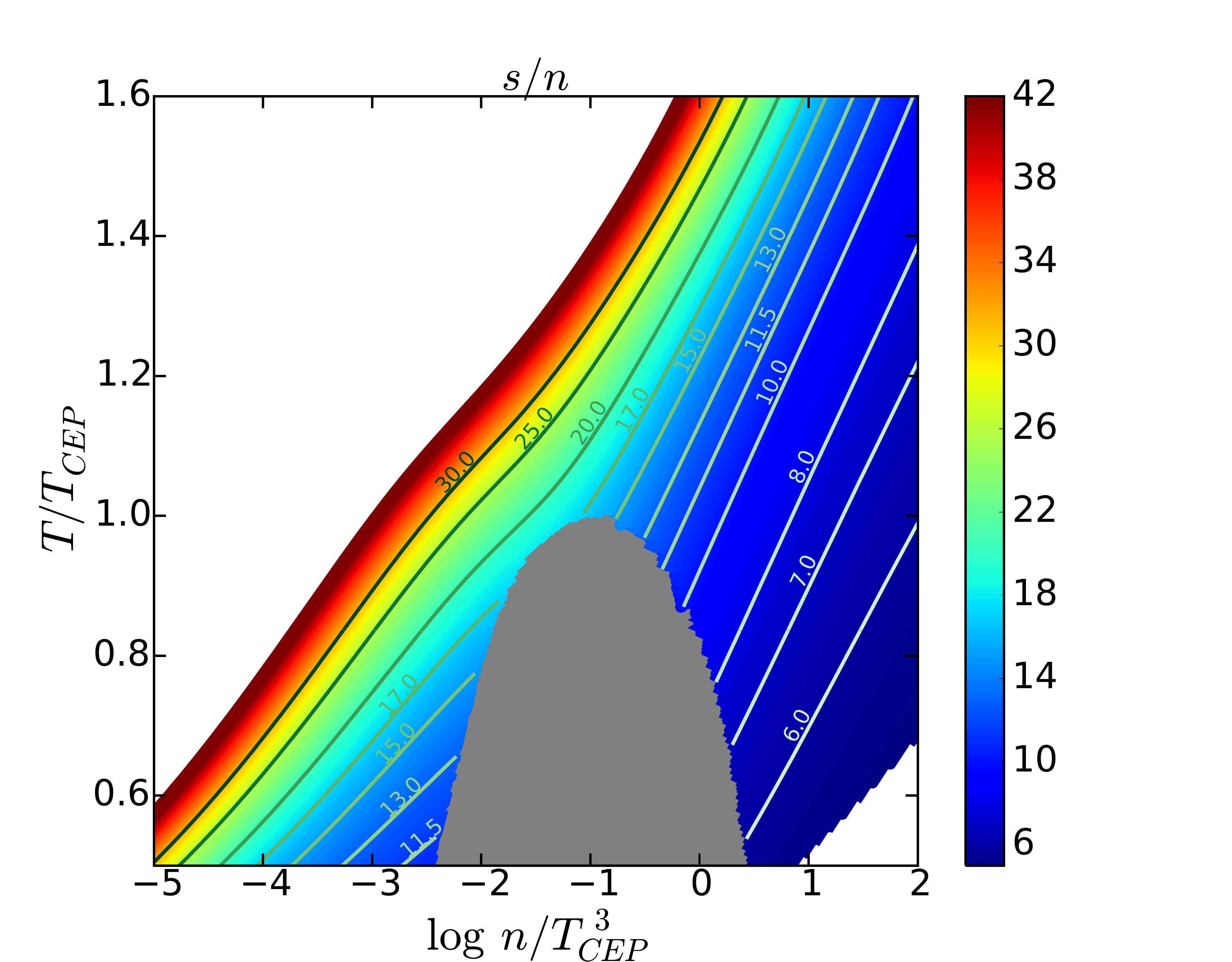}
 \includegraphics[width=0.495\textwidth]{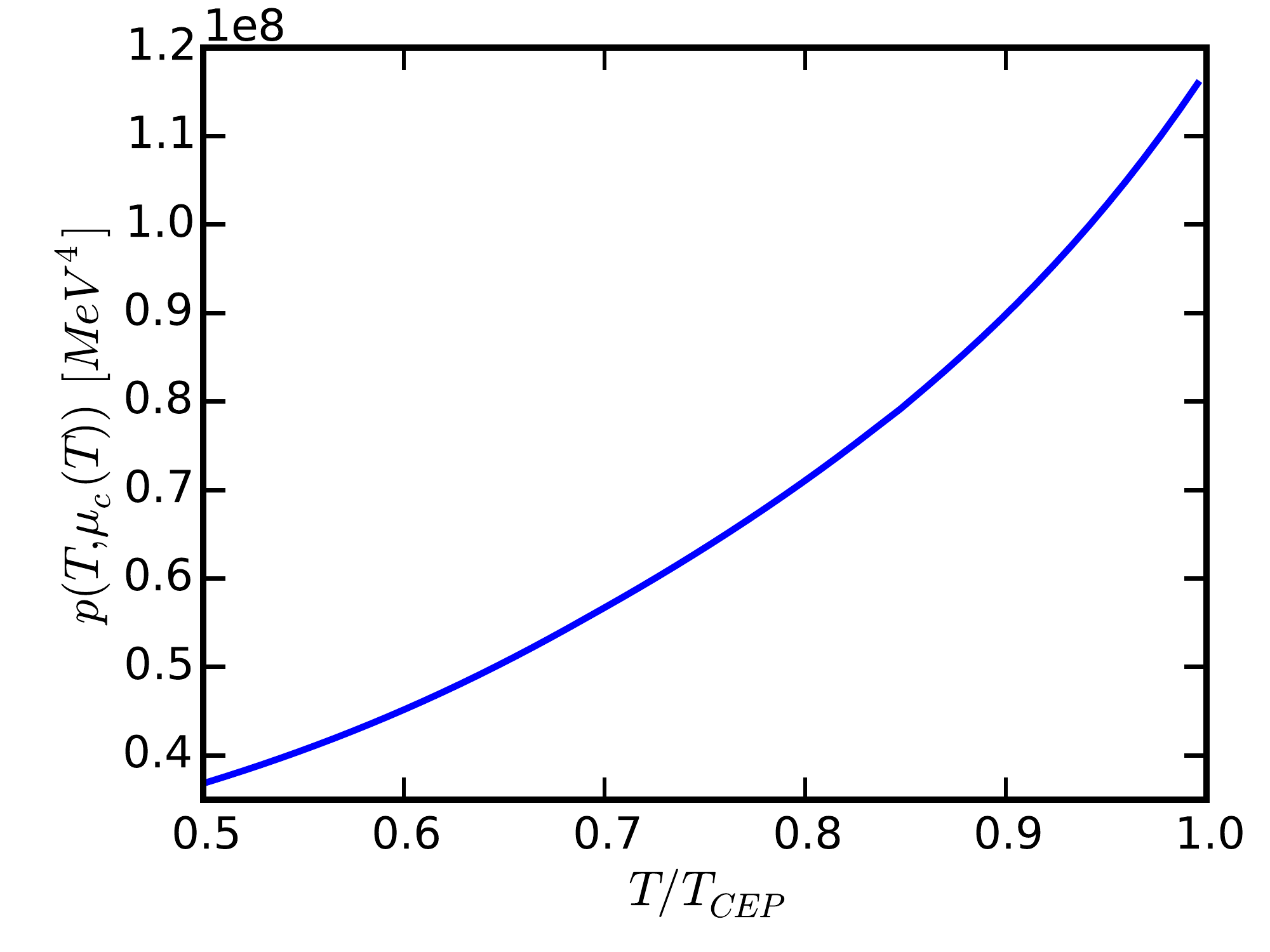}
 \caption{Left panel: Contour plot of entropy-to-baryon ratio over the $T/T_{CEP}-\log(n/T_{CEP}^3)$ plane for the updated holographic EMd model. The coexistence region is shown in grey. White regions indicate areas that are beyond the range of the colorbar.
 Right panel: Critical pressure $p_c = p(T, \mu_c(T))$ for the updated holographic EMd model.}
 \label{fig:pressure_sn}
\end{figure}

The FOPT curve shows up as kinky  behavior of the pressure and jumpy behavior of the entropy density, baryon density and entropy-to-baryon ratio (cf.\ corresponding panels in Fig.\,\ref{fig:PD}).

The contour curves in the pressure panel in Fig.\,\ref{fig:PD} are scaled isobars, $p/T^4 = \text{const}$.
The pressure increases in $\mu$-direction (e.g.\ at constant $T$).
The FOPT curve is steeper than neighboring isobars, as characteristic for the GL FOPT. 
This implies that the critical pressure $p_c(T) = p(T, \mu_c(T))$ increases with 
temperature, see right panel in Fig.\,\ref{fig:pressure_sn}. 
The information on both entropy density and 
baryon density (see top right and bottom left panel in Fig.\,\ref{fig:PD}) can be combined to the contour plot of
constant entropy per baryon, see bottom right panel in Fig.\,\ref{fig:PD}. The resulting contour curves 
are isentropes, i.e.\  paths of gas or fluid elements during an adiabatic expansion 
(collapse) stage in heavy-ion collisions (stellar core collapse). 
The scaled pressure, entropy density and baryon density are pushed towards higher values with increasing chemical potential, whereas the entropy-to-baryon ratio is decreasing. 
For $\mu\ge\mu_{CEP}$ the scaled entropy density, scaled baryon density and entropy-to-baryon ratio jump across the FOPT.

Comparing the entropy-to-baryon ratios on both sides of a point on the FOPT 
curve evidences $s/n\vert_{-} > s/n\vert_{+}$, where the label $ - (+)$ means 
approaching the FOPT curve from left (right). In line with the above mentioned
Clausius-Clapeyron relation 
$dp_c(T)/dT = \big(\frac{s}{n}\big\vert_{-} - \frac{s}{n}\big\vert_{+}\big)\big(\frac{1}{n}\big\vert_{-} - \frac{1}{n}\big\vert_{+}\big)$, 
this implies in turn that the curve $p(T, \mu_c(T))$ 
as a function of $T$ has positive slope, see Fig.\,\ref{fig:pressure_sn} (right panel), as typical for the GL transition. 
Isentropes meeting the FOPT curve are ``incoming" on the 
deconfined/dense ($+$) side and ``outgoing" on the confined/dilute ($-$) side. 
Figure\,\ref{fig:pressure_sn} (left panel) illustrates this behavior in the $T-\log n$ plane. 
The two-phase coexistence region is depicted by a grey region, 
where the isentropes (not displayed) are to be constructed by the lever rule. 
The panel exposes the shape of the isentropes as paths of adiabatically expanding and cooling pieces of matter: In the updated EMd model, isentropes enter \textit{and} leave the coexistence region.
This is in contrast to \cite{DeWolfe:2010he,DeWolfe:2011ts} where only incoming isentropes can be found.
According to the nomenclature of \cite{Wunderlich:2016aed}, the updated EMd model is classified as type IA and represents a GL phase transition.

{\color{change}In addition to the second-order quark number susceptibility at $\mu=0$, Fig.\,\ref{fig:chis} exhibits also $\chi_2/T^2$
for finite $\mu$ (green and red curve in the left panel).
With increasing chemical potential, $\chi_2/T^2$ is pushed towards larger values. 
A maximum is evolving, which transforms into a divergence at the CEP.}

The updated holographic EMd model is based on a fit to the recent lattice data for $\mu=0$. 
Since no lattice results are available for low-temperatures and the critical point is located in a temperature range where the lattice data of $\chi_{2,4}$ just start, we estimate this uncertainty by {\color{change}allowing parameter variations of the dilaton potential and gauge kinetic function within the above mentioned uncertainties} and assuming different generic low-temperature asymptotics for the second-order quark number susceptibility {\color{change}and dilaton potential (i.e.\ continuing the data exhibited in Fig.\,\ref{fig:chis} to zero with different slopes)}. 
The different types just slightly vary $T_{CEP}$ in the order of 
\unit[5]{MeV} and $\mu_{CEP}$ in the order of \unit[50]{MeV}, corresponding to a relative uncertainty of approximately $5\,\%$ {\color{change}and $8\,\%$} respectively.

\section{Summary}
\label{sec:summary}

In summary we explore here the phase structure of the holographic 
Einstein-Maxwell-dilaton (EMd) model \cite{DeWolfe:2010he,DeWolfe:2011ts} adjusted now at 2 + 1 flavor lattice QCD data 
with physical quark masses at $\mu = 0$. 
The EMd model has a first-order phase transition (FOPT) curve
setting in at a (critical) endpoint (CEP) with coordinates \TCEP, 
{\color{change}\MUCEP}. 
By considering {\color{change}parameter variations and} different low-temperature asymptotics for the second order quark number susceptibility {\color{change}and equation of state that take the uncertainties of the lattice data into account}, we estimate the relative uncertainty of our result for the critical point position in the {\color{change}range up to $8\,\%$}. 
We emphasize that these values are consistent with recent lattice estimates in \cite{Bazavov:2017dus} {\color{change}and the covered range of the phase diagram in \cite{Gunther:2016vcp}}. 

The FOPT curve continues from the CEP towards the
$T = 0$ axis. Recalling the general remarks in \cite{Rougemont:2015wca,Rougemont:2015ona} we refrain from 
analyzing the region of small temperatures 
(i.e.\ we leave the quantum phase transition for separate consideration), 
which however is relevant w.r.t.\ 
neutron (quark) stars and particular scenarios for core-collapse supernova explosions. 
The EMd model does not include explicitly such QCD relevant aspects as chiral 
symmetry or confinement. Instead, it accounts implicitly for these fundamental 
notions by the adjustment at QCD results. 
{\color{change2}In particular, our holographic bottom-up approach does not explicitly include the physics of the chiral condensate.
Effects as chiral symmetry breaking and restoration can be studied by taking into account the backreaction of flavored branes in the bulk.
A holographic model for such important considerations was put forward in \cite{Jarvinen:2011qe} (with further developments in \cite{Alho:2012mh,Arean:2013tja,Alho:2013hsa,Jarvinen:2015ofa,Gursoy:2016ofp,Gursoy:2017wzz}) for the Veneziano limit of QCD. 
Instead of accommodating all wanted physics solely in the dilaton potential, the gluon and quark degrees of freedom are shared by the glue sector and a coupled flavor sector being dual to the real-scalar dilaton and a complex scalar for the quark condensate $\langle\bar q q\rangle$.  
This procedure of adding flavors is appropriate to address also magnetic field effects and enjoys a qualitative agreement with lattice results.
The relationship of genuinely non-perturbative properties such as color confinement and spontaneous chiral symmetry breaking poses another issue in this context.}

The resulting phase diagram {\color{change2}in the present (minimalistic) approach with
a dilaton solely} resembles in many aspects the gas-liquid phase transition.
For instance, the critical pressure 
increases with temperature, contrary to expectations of the hadron-quark transition.
In the updated EMd model, isentropes are incoming from the dense phase, enter the coexistence region, run through and leave the critical curve at lower temperature. 
The updated EMd model exhibits a graceful exit into the pure low-temperature and low-density phase.

We emphasize the need to supplement (model) phase diagrams by information 
on isobars or isentropes, for instance, for having access to physics implications.

{\color{change}Recently, other ans\"atze for the dilaton potential and gauge kinetic function were presented in \cite{Critelli:2017oub}, which result in CEP coordinates 
$T_{CEP}=\unit[(89\pm11)]{MeV}$ and $\mu_{CEP}=\unit[(723\pm36)]{MeV}$, only marginally consistent with our result $T_{CEP}=\unit[(112\pm5)]{MeV}$ and $\mu_{CEP}=\unit[(612\pm50)]{MeV}$,
where the uncertainties refer to the above quoted $8\,\%$.
Since both setups allow an equally good description of the available lattice data for the equation of state and susceptibilities, we conclude that the underlying holographic model is rather sensitive to both the input data 
and internal parametrization, which affect the CEP position and phase structure.
It is in particular the lacking precision lattice data at $T \approx \unit[100]{MeV}$ which seems to hamper a unique determination of CEP coordinates.}  \\

Acknowledgements: Enlightening conversations with J.\ Noronha on holographic 
models are gratefully acknowledged. 
We thank J.\ Randrup and V.\ Koch for discussions on phase transitions and S.\ Borsanyi 
for supplying data of susceptibilities shown in \cite{Bellwied:2015lba}. 
Options for the hadron-quark transition have been discussed with B.\ Friman some time ago.

\section*{References}

\bibliographystyle{jk_ref_layout_noTitle}
\bibliography{literature}




\end{document}